\documentstyle[11pt,nature,psfig]{article}
\lefthead{Srianand, Petitjean and Ledoux}
\righthead{}

\begin{document}
\author{\raggedleft To appear in Nature}

\title{The Cosmic Microwave Background temperature at a 
redshift of 2.33771\\}

\author{
\raggedright R. Srianand\altaffilmark{1},
P. Petitjean\altaffilmark{2,3},
C. Ledoux\altaffilmark{4}
}
\affil{\raggedright
1) IUCAA, Post Bag 4, Ganeshkhind, Pune 411 007, India\\
2) Institut d'Astrophysique de Paris -- CNRS, 98bis Boulevard Arago, F-75014 Paris, France\\
3) CNRS 173 -- DAEC, Observatoire de Paris-Meudon, F-92195 Meudon Cedex, France \\
4) European Southern Observatory, Karl Schwarzschild Strasse 2, D-85748 Garching bei M\"unchen, Germany
}
\clearpage
\newpage
\par\smallskip\noindent
{\bf
The Cosmic Microwave Background radiation is a fundamental prediction
of Hot Big Bang cosmology. The temperature of its black-body spectrum
has been measured at the present time, $T_{\rm CMBR,
0}$~=~2.726$\pm$0.010~K, and is predicted to have been higher in the
past.  At earlier time, the temperature can be measured, in principle,
using the excitation of atomic fine structure levels by the radiation
field.  All previous measurements however give only upper limits as
they assume that no other significant source of excitation is present.
Here we report the detection of absorption from the first {\sl and}
second fine-structure levels of neutral carbon atoms in an isolated
remote cloud at a redshift of 2.33771. In addition, the unusual
detection of molecular hydrogen in several rotational levels and the
presence of ionized carbon in its excited fine structure level make the
absorption system unique to constrain, directly from observation, the
different excitation processes at play.  It is shown for the first time
that the cosmic radiation  was warmer in the past.  We find
6.0~$<$~$T_{\rm CMBR}$~$<$~14~K at $z$~=~2.33771 when 9.1~K is expected
in the Hot Big Bang cosmology.
}
\par\bigskip\noindent
One of the firm predictions of standard Big Bang model is the
existence of relic radiation from the hot phase the Universe
has experienced at early times$^{\cite{ah1948}}$.
The cosmic microwave background radiation (CMBR) has been discovered 
serendipitously by Penzias \& Wilson$^{\cite{penzias}}$ in 1964.
The fact that its spectrum follows with remarkable
precision a Planckian distribution over several decades in frequency
is a strong argument in favor of the Hot Big Bang cosmology. 
However the presence of the radiation at earlier time 
has never been proven directly. The temperature of its black-body
spectrum is predicted to increase linearly with redshift $T_{\rm
CMBR}$($z$)~=~$T_{\rm CMBR}$(0)$\times$(1~+~$z$) and its local value has
been determined very accurately by the Cosmic Background Explorer
(COBE) $T_{\rm CMBR}$(0)~=~2.726$\pm$0.010~K$^{\cite{mat}}$.
Detecting the  presence of relic radiation at earlier epochs and confirming 
the well defined temperature evolution is therefore a crucial
test for cosmology.
To perform this test, one can use the possibility that
excited fine structure levels of the ground-state of atomic species can
be partly populated by the Cosmic Microwave Background Radiation when
the energy separation of the levels is similar to the energy at the
peak of the radiation energy distribution. 
\par\noindent
The relative populations of excited levels can be measured from the 
absorption lines seen in the spectrum of distant quasars and used
therefore to constrain $T_{\rm CMBR}$ at high 
redshift$^{\cite{bah73}, \cite{meyer},\cite{son1},
\cite{son2}, \cite{Lu},\cite{ge},\cite{roth}}$.  The expected values
make neutral carbon, C$^0$, particularly well suited for this purpose.
The ground term is split into three levels ($J$~=~0, 1, 2) with
$J$~=~0--1 and $J$~=~1--2 energy separations $kT$ corresponding to,
respectively, $T$~=~23.6 and 38.9~K.  However the fine structure levels
can also be excited by collisions (mostly with electrons and hydrogen)
and by UV pumping and following cascades.  The ionization potential of
neutral carbon, 11~eV, is below the ionization potential of hydrogen,
13.6~eV, and C$^0$ can only be seen in dense, neutral and highly
shielded gas$^{\cite{ge}}$. Therefore, excitation by collisions cannot
be neglected. \par\noindent To constrain $T_{\rm CMBR}$, the kinetic
temperature, the particle density of the gas and the local UV radiation
field must be known. As it is very difficult to disentangle these various
excitation processes, all measurements up to now have lead to upper
limits on $T_{\rm CMBR}$ only.
\par\noindent
Our measurement has been obtained in a unique absorption system where
absorption lines of neutral carbon in the three fine-structure level of
the ground term, C$^0$, C$^{0*}$, C$^{0**}$ are observed together with
absorption lines of singly ionized carbon in its excited fine-structure
level, C$^{+*}$, and absorption lines of molecular hydrogen (H$_2$) in the 
$J$~=~0 to 5 rotational levels. The population and depopulation of the first 
excited rotational level of H$_2$ ($J$~=~1) from and to the ground state
($J$~=~0) is controlled by thermal collisions. Therefore the excitation
temperature $T_{01}$ is approximately equal to the kinetic temperature.
The fine structure upper level of the C$^+$ ground-state doublet is
mostly populated by collisions and depopulated by radiative decay.
Therefore, once the temperature is known, the particle density can be
derived from the C$^{+*}$/C$^+$ ratio. Finally, the UV radiation flux
can be constrained from the populations of the J~=~4 and 5 H$_2$
rotational levels.
\par\medskip\noindent
\noindent{\large \bf Observations}
\par\medskip\noindent
We have used the Ultra-violet and Visible Echelle Spectrograph
(UVES)$^{\cite{UVES}}$ mounted on the ESO KUEYEN 8.2~m telescope at the
Paranal observatory on April 5 and 7, 2000 to obtain a high-spectral
resolution spectrum of the $z_{\rm em}$~=~2.57 and $m_{\rm V}$~=~18.4
quasar PKS~1232+0815.  Standard settings have been used in both arms of
the spectrograph.  Wavelength ranges were 3290--4519~\AA~ in the blue;
4623--5594 and 5670--6652~\AA~ for the red chips.  The slit width was
1~arcsec and the CCDs were binned 2$\times$2 resulting in a resolution,
$\Delta \lambda/\lambda$, of $\sim$45000. The exposure time was 3 hours
in seeing conditions better than 0.8~arcsec full width at half
maximum.  The data were reduced using the UVES pipeline in an interactive
mode. The pipeline is  a set of
procedures implemented in a dedicated context of MIDAS, the ESO data
reduction package. The main characteristics of it is to
perform a precise inter-order background subtraction for science frames
and master flat-fields, and to allow for an optimal extraction of the
object signal rejecting cosmic ray impacts and performing
sky-subtraction at the same time.  The reduction is checked step by
step. Wavelengths are corrected to vacuum-heliocentric values and
individual 1D spectra are combined together.  This resulted in a S/N
ratio per pixel of 10 around 3700~\AA~ and  20 around 6000~\AA.  
Typical errors in the wavelength
calibration is $\sim$0.5~km~s$^{-1}$.
\par\medskip\noindent
\noindent{\large \bf Fit to the lines, metallicity and molecular content}
\par
\begin{figure}
\centerline{\vbox{
\psfig{figure=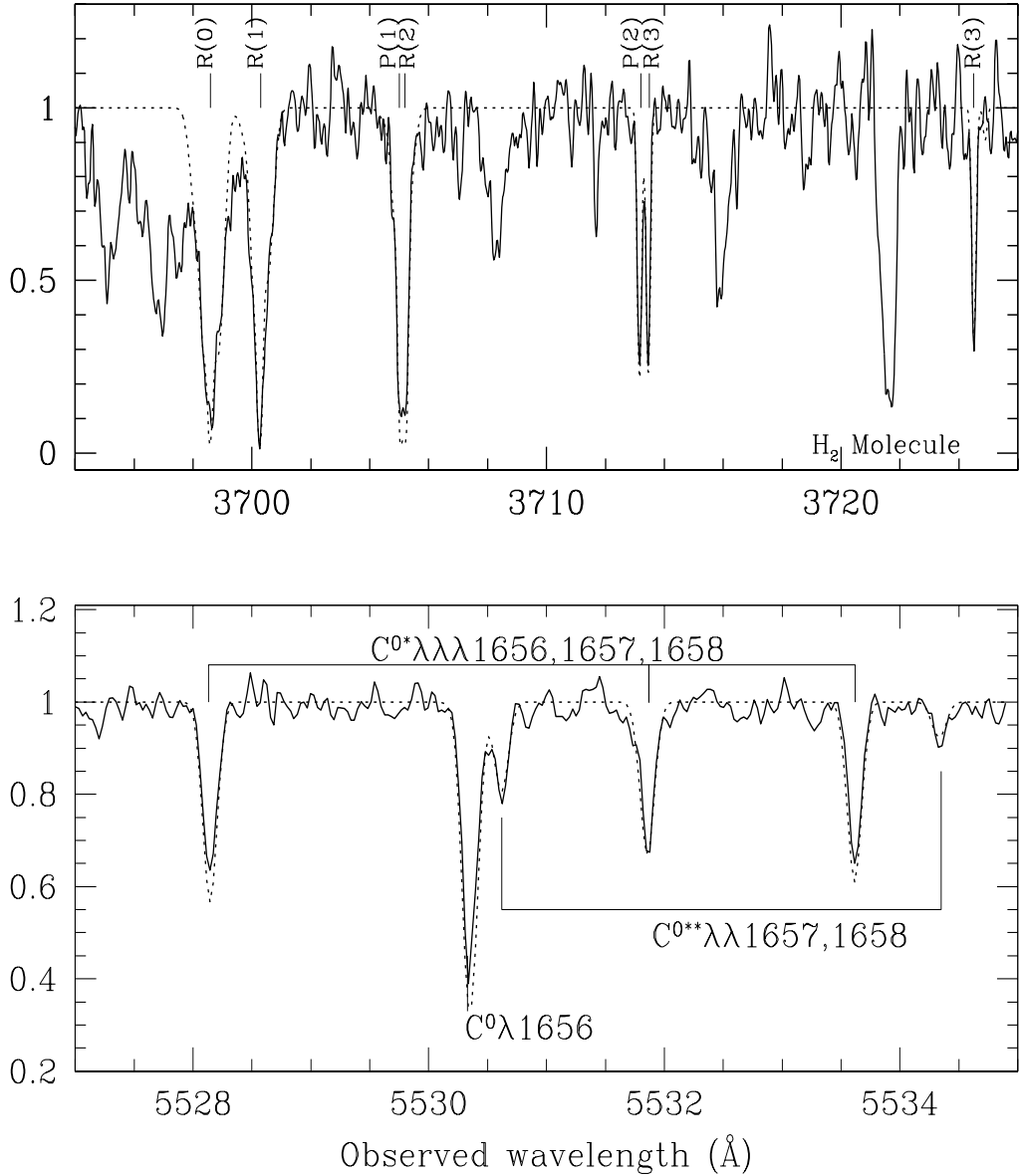,height=17.cm,width=15.cm,angle=0}
}}
\figcaption[cmbfig1.ps]{{\bf A sample of H$_2$ and C$^0$ absorption lines
at $z_{\rm abs}$~=~2.33771:}
Portions of the normalized spectrum of the quasar PKS~1232+0815
taken with the Ultra-violet and Visible Echelle Spectrograph mounted
on the 8.2~m KUEYEN telescope of the European Southern Observatory
on the Paranal mountain in Chile. 
{\sl Upper panel:} A selection of H$_2$ absorption lines
from the $J$~=~0, 1, 2 and 3 rotational levels from the 
v~=~0--1 Lyman band. 
The model fit with parameters given in Table~2 is overplotted to the data as a 
dashed line.  
{\sl Lower panel:} Detection of absorption lines from C$^0$, 
C$^{0*}$ and C$^{0**}$ at $z_{\rm abs}$~=~2.33771 
in the damped Lyman-$\alpha$ system. 
The model fit with parameters given in Table~1 is over-plotted to the data as a 
dashed line.\label{CI} }
\end{figure}

\begin{figure}
\centerline{\vbox{
\psfig{figure=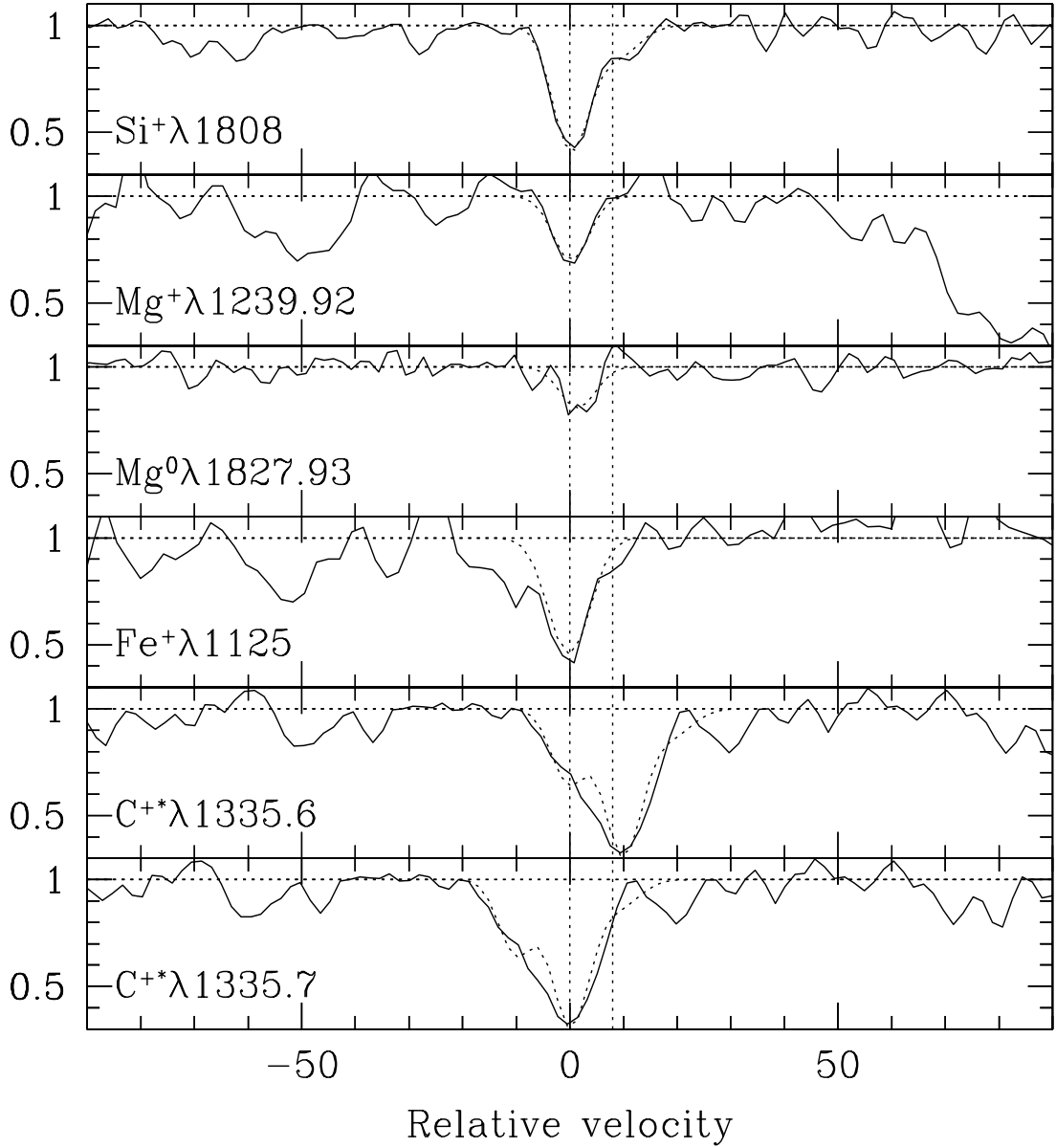,height=17.cm,width=16.cm,angle=0}
}}
\figcaption[cmbfig2.ps]{{\bf A sample of heavy element absorption lines at
$z_{\rm abs}$ = 2.33771:}
Absorption profiles of a few transitions (indicated on each panel)
from the damped Lyman-$\alpha$ system toward PKS~1232+0815.
The normalized flux is given on a velocity scale with origin at $z$~=~2.33771.
The C~$^{+*}$ profile is slightly broader than the other lines.
The C$^{+*}$ line is a blend of C$^{+*}\lambda1335.71$ and
C$^{+*}\lambda1335.66$.
Fitted models are over-plotted as dashed lines.
The fit is performed over all the absorption lines available in the
spectrum.
Our best fitted column density of $10^{14}$ cm$^{-2}$ for C$^{+*}$
overpredicts the absorption at $\lambda$1335.6.
\label{metals}}
\end{figure}

\medskip\noindent
This is the first time C$^0$, C$^{0*}$ {\sl and} C$^{0**}$ absorption
lines are detected at high redshift (see Fig.~\ref{CI}). The detection
of these three species at $z_{\rm abs}$~=~2.33771 is confirmed by the
presence of several transitions. C$^0$ absorption lines with rest
wavelengths 1139.79, 1157.91, 1260.73, 1560.31 and 1656.92~\AA~,
C$^{0*}$ absorption lines with rest wavelengths 1194.40, 1279.89,
1329.09, 1329.10, 1329.12, 1560.68, 1560.71, 1656.27, 1657.38~\AA~ and
1657.91 and C$^{0**}$ lines at 1657.00 and 1658.12~\AA~ are clearly
detected and are free from any blending with absorption due to other
systems. We use the oscillator strengths given by Welty et
al.$^{\cite{welty}}$ and standard Voigt profile fitting to determine
the amount of matter lying on the line of sight and characterized by
column densities, $N$, and the Doppler parameters, $b$~=~$\sqrt{b_{\rm
th}^2+b_{\rm turb}^2}$ where $b_{\rm th}$~=~${\sqrt{2KT/m}}$ is the
thermal broadening due to the temperature of the gas, with $T$ the
kinetic temperature, $K$ the Boltzman constant and $m$ the mass of the
particle and $b_{\rm turb}$ is the turbulent broadening due to
macroscopic motions of the gas. It can be seen in Figs.~\ref{CI} and
\ref{metals} that the system is dominated by a single and well defined
component. We consistently impose the same $b$ value to this component
for all species and obtain a best fit for $b$~= 1.7$\pm$0.1 kms$^{-1}$.
This value is small compared to the spectral resolution of the data but
is ascertained by the relative optical depths in the numerous lines
with very different oscillator strengths.  Results are given in Table~1
and the fit to a portion of the spectrum is shown in the bottom panel
of Fig.~\ref{CI} and in Fig.~\ref{metals}.
\par\noindent
The neutral hydrogen column density in the system,
log~$N$(H$^0$)~=~20.90$\pm$0.10, has been obtained by fitting the
damped Lyman-$\alpha$ line. At such a high column
density, hydrogen is mostly neutral and the elements of interest
here, iron, magnesium, silicon and carbon are mainly in the singly ionized
state.  Column densities are derived from simultaneous Voigt profile
fitting of all available absorption lines using the same Doppler 
parameter for
all species and the same column density for each species.  We
concentrate however on weak lines as Fe$^+$$\lambda\lambda
\lambda$1125,1608,1611, Si$^+$$\lambda$1808 or Mg$^+$$\lambda$1239
because they are optically thin. Indeed in this case the strength of the line 
does not depend on the Doppler parameter and directly gives the column 
density. Profiles of a few transitions are shown on Fig.~\ref{metals} and
results are given in Table~1.  The $\alpha$-chain elements magnesium
and silicon have similar abundances, $Z({\rm X})$~=~$N$(X)/$N$(H),
relative to solar within the measurement uncertainties,
log~$Z/Z_{\odot}$~$\sim$~$-$1.2. However we note that, relative to
what is seen in the sun, iron is underabundant by about a factor of 5
compared to silicon and nickel is slightly underabundant compared to
iron.  If these differences are due to iron and nickel being depleted
into dust-grains, the depletion is of the same order as what is
observed in warm halo gas in our Galaxy$^{\cite{savage}}$.  All this
suggests that the system is very much like typical damped
Lyman-$\alpha$ systems, with metallicity about one tenth of solar and
small depletion of iron type element into dust$^{\cite{Lu},\cite{pettini}}$.

\par\noindent
Our echelle data confirms the presence of molecular hydrogen first
detected by Ge and Bechtold$^{\cite{ge1}}$.  At the resolution of the
spectrum, most of the H$_2$ lines are free from blending. The wide
wavelength coverage of the spectrum implies that absorption lines from
various Lyman-bands are seen.  Column densities of H$_2$ in different
rotational $J$ levels are obtained by simultaneously fitting various
Lyman-bands which are free from contamination due to intervening
Lyman-$\alpha$ absorption from the diffuse intergalactic medium.
Absorption profiles corresponding to transitions from $J$~$>$~1
rotational levels are consistent with a single component at the same
redshift $z_{\rm abs}$~=~2.33771 as the C$^0$ lines. Lines from the
$J=0$ and 1 levels are broader and most of the absorption features have
similar profiles suggesting the presence of several components. We
obtain a best fit to these absorption lines with four components.
Results of Voigt profile fitting are given in Table~2.  The total H$_2$
column density is $N$(H$_2$) = 1.52$\times10^{17}$~cm$^{-2}$ and the
molecular fraction is $f$ = 2$N$(H$_2$)/(2$N$(H$_2$)+$N$(H$^0$)) =
3.8$\times$10$^{-4}$ which is about two orders of magnitude less than
what was derived from low dispersion data.
\par\medskip\noindent
\noindent{\large \bf Physical conditions in the gas
}
\par\medskip\noindent
In molecular clouds, the kinetic temperature of the gas is given by the
excitation temperature, $T_{\rm 01}$, measured between the $J$~=~0 and
1 levels.  For the components at $z_{\rm abs}$~=~2.33735, 2.33754 and
2.33793 we find $T_{01}$~= $74\pm7$, 64$\pm7$ and 66$\pm10$~K
respectively.  None of these components show detectable absorption due
to C$^0$. The characteristics of the gas are very similar to what has
been measured recently in diffuse gas in the halo of our
Galaxy$^{\cite{shull}}$.
For the $z_{\rm abs}$~=~2.33771 cloud, $T_{\rm
kin}$~=~$T_{01}$~=~185$\pm{100}$~K.  The excitation temperature for
other levels, $J$~=~2 to 5, is close to $<T_{ex}>~=~400$~K.  This means
that processes other  than collisions are at play in determining the
relative populations of the different levels.  Indeed, the  $J~=~4$ and
$~5$ levels can be populated by cascades following UV-pumping and H$_2$
formation. We can evaluate the UV-pumping rate from $J$~=~0 to $J$~=~4,
$\beta$(0), using the simple model of Jura$^{\cite{jura}}$ and assuming
that molecules are formed on dust grains with formation rate $R$,
\begin{equation}
p_{4,0}\beta(0)n({\rm H_2}, J=0) + 0.19Rn(H)n
 ~=~A(4\rightarrow 2)n({\rm H_2}, J= 4)\nonumber 
\end{equation}
where $n = n(H)+2n({\rm H_2})$, $p_{4,0}~=~0.26$ is the pumping
efficiency from level $J=0$ to level $J=4$.  We use the spontaneous
transition probabilities $~A(4\rightarrow
2)~=~2.8\times10^{-9}$~s$^{-1}$ as given by Dalgarno \&
Wright$^{\cite{dalgarno}}$.  Scaling the formation rate of H$_2$ in our
Galaxy ($R\sim3\times10^{-17}~{\rm s^{-1}}~{\rm cm}^{3}$) with the dust
to hydrogen ratio measured in the system ($\le10^{-1.3}$ the Galactic one) 
it is easily
seen that populating the $J=4$ level following formation of a molecule
on dust grain is negligible for densities less than
5$\times10^3$~cm$^{-3}$. We estimate the density from the excitation of
C$^+$ to be at least a factor of hundred less than this value (see below).  
Moreover, for such large densities, the $J=0$ and $J=2$ levels as well as the
$J=1$ and $J=3$ levels should be in thermal equilibrium which is not
the case$^{\cite{sri}}$.  We therefore can neglect populating of $J=4$
and $ 5$ levels by cascades after the formation of a molecule.
We finally derive from Equation~1 that the photo-absorption rate in the 
Lyman and Werner bands in the surface of the cloud is of the order of
2$\times$10$^{-10}$~s$^{-1}$ which is quite modest and similar to that
observed along sight-lines in our Galaxy$^{\cite{jura}}$.
\par\medskip\noindent

The excited level of the C$^+$ ground-state term is populated by
collisions with electrons and hydrogen atoms$^{\cite{bah68}}$.
Electrons cannot be neglected because the corresponding collisional
cross-section is large. To estimate the electron density we use the
ratio $N$(Mg$^+$)/$N$(Mg$^0$)~=~178 which is fairly well determined
from the detection of weak lines Mg$^0$$\lambda\lambda$1737,1827 and
Mg$^+$$\lambda$1239.  As we have shown above that the ionizing flux in
the cloud is similar to that observed in our Galaxy we can estimate
the electron density by equating the Galactic ionizing rate of Mg$^0$
to the recombination rate of Mg$^+$$^{\cite{welty}}$.  We find $n_{\rm
e}$~$\sim$~0.02~cm$^{-3}$. Note that the electronic density cannot be
much smaller than this value because, from the excitation of the $J=4$ and 5
H$_2$ levels, we know that the ionizing flux is not smaller than the
Galactic value. \par\noindent Then, the hydrogen density can be derived
from the $N$(C$^{+*}$)/$N$(C$^+$) ratio. Unfortunately, the
C$^+$$\lambda$1334 absorption line is saturated and cannot be used
directly to infer $N$(C$^+$).  We therefore consider that the carbon
metallicity can be derived from the silicon metallicity with some
correction. Indeed, it is known that the metallicity of the
$\alpha$-chain elements is enhanced compared to carbon by a factor of
about two$^{\cite{timmes}}$ when the mean metallicity is low (i.e.
$\le-1.0$).  To be conservative, we consider that $Z$(C)~=~$Z$(Si)/2
{\sl and} that silicon is not depleted into dust. Note that both
assumptions maximize the C$^{+*}$/C$^+$ ratio and, as a consequence,
maximize the derived hydrogen density, $n_{\sc H}$. In addition, the
very small C$^0$/C$^+$ value shows that the ionization correction is
negligible. Note also that absorptions due to Si$^{3+}$ and C$^{3+}$
are unusually weak in this damped system.  The fit of the
C$^{+*}$$\lambda\lambda$1335.6,1335.7 lines is shown in
Fig.~{\ref{metals}}. As can be seen from this figure the observed
C$^{+*}$ profile suggests the presence of possible extra components.
Our measured column density is therefore an upper limit which will turn
into an upper limit on the hydrogen density.  It must be noted that the
strongest constraint comes from the 1335.6 line in the blue wing of the
profile.
\par\noindent

\begin{figure}
\centerline{\vbox{
\psfig{figure=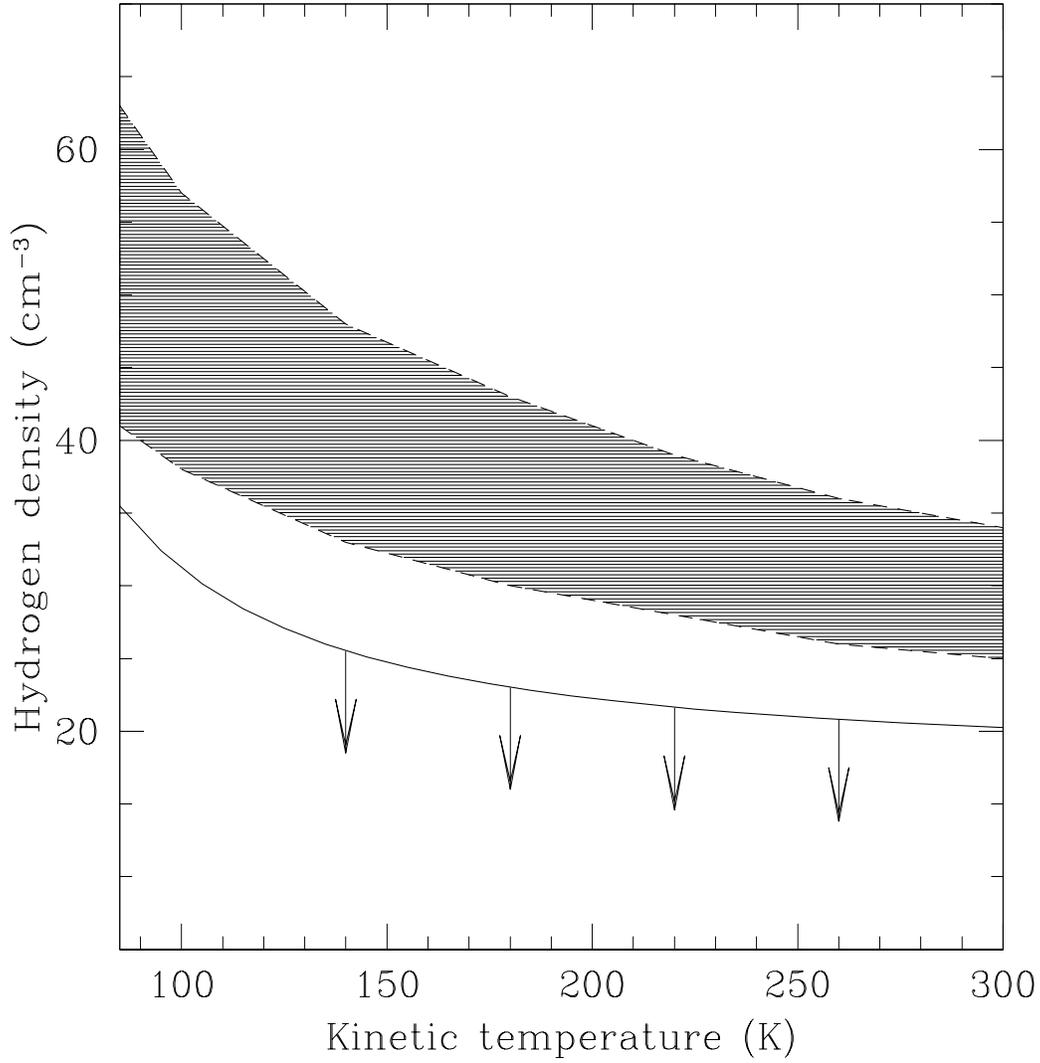,height=15.cm,width=14.cm,angle=0}
}}
\figcaption[cmbfig3.ps]{ {\bf The hydrogen density as a function of kinetic temperature
in the $z_{\rm abs}$=2.33771 cloud:} 
The continuous curve with downward arrows shows the strict upper limit on the
hydrogen density estimated from the fine-structure excitation of
C$^+$.  
The shaded area gives the 2$\sigma$ range for the hydrogen density
required to explain the population ratios in the fine-structure 
levels of C$^0$ assuming there is no cosmic background radiation.
All other excitation processes (collisions by
electrons and hydrogen atoms, UV pumping) have been taken into account.
It is apparent that the observed density is too small to explain
the C$^0$ excitation demonstrating that the Cosmic Microwave Background
exists at $z$~=~2.33771.
\label{den}}
\end{figure}

Results on the determination of $n_{\rm H}$ are presented in
Figure~{\ref{den}}. For a kinetic temperature in the range
85~$<$~$T$~$<$285~K, as suggested from the H$_2$ absorption, we find
20~$<$~$n_{\rm H}$~$<$~35~cm$^{-3}$ (solid curve on Fig.~{\ref{den}}).
All assumptions above maximize the hydrogen density which can
therefore be considered for each temperature as a conservative upper
limit.
\par\noindent

\noindent{\large \bf Cosmic Microwave Background radiation temperature}
\par\noindent
The fine structure levels of C$^0$ can be populated by several
processes, mainly collisions with hydrogen atoms and electrons and
pumping due to the local UV-radiation and to the CMBR.  The different
contributions can be estimated once the temperature and the particle
density are known. We have derived above 85~$<$~$T$~$<$~285~K,
20~$<$~$n_{\rm H}$~$<$~35~cm$^{-3}$ and $n_{\rm e}$~/$n_{\rm
H}$~$\sim$~0.001.  \par\noindent Following Meyer et
al.$^{\cite{meyer}}$ and using cross-sections given by Nussbaumer \&
Rusca$^{\cite{nussbaumer}}$ and Launay \& Roueff$^{\cite{launay}}$, we
investigate the fine-structure excitation of C$^0$ keeping the
particle density and the temperature within the ranges determined in the
previous Sections. For a
radiation field similar to that of the Milky-way$^{\cite{jenkins}}$,
the UV pumping rate from the ground state to the excited states of
C$^0$ is 7.55$\times10^{-10}~{\rm s}^{-1}$.  The corresponding value
for the hydrogen collisional rate in the density and temperature ranges
considered here is of the order of $\sim 1.2-1.4 \times 10^{-8}~{\rm
s}^{-1}$. Thus the contribution due to UV pumping is negligible.
Keenan et al.$^{\cite{keenan}}$ noticed that collisions with electrons
are unimportant for kinetic temperatures in the range 100--500 K and a
ratio of electron density to hydrogen density less than 0.01.
Collisions with electrons can therefore be neglected.
\par\noindent
The observed 2$\sigma$ range for $N$(C$^{0*}$)/$N$(C$^0$) and
$N$(C$^{0**}$)/$N$(C$^0$) are, respectively, 0.28$-$0.48 and
0.020$-$0.060.  Assuming collision by hydrogen atoms is the only
process at work, we estimate the range of hydrogen density needed to
explain the C$^0$ populations.  This density range is shown as a shaded
region in Fig.~\ref{den}.  From this figure it is apparent that the
observed upper limit on the hydrogen density derived above is not
sufficient to populate the excited fine-structure levels of C$^0$ and
that an additional source of excitation is needed.  The only process we are
left with is direct pumping due to photons from the relic background
radiation.
\par
\begin{figure}
\centerline{\vbox{
\psfig{figure=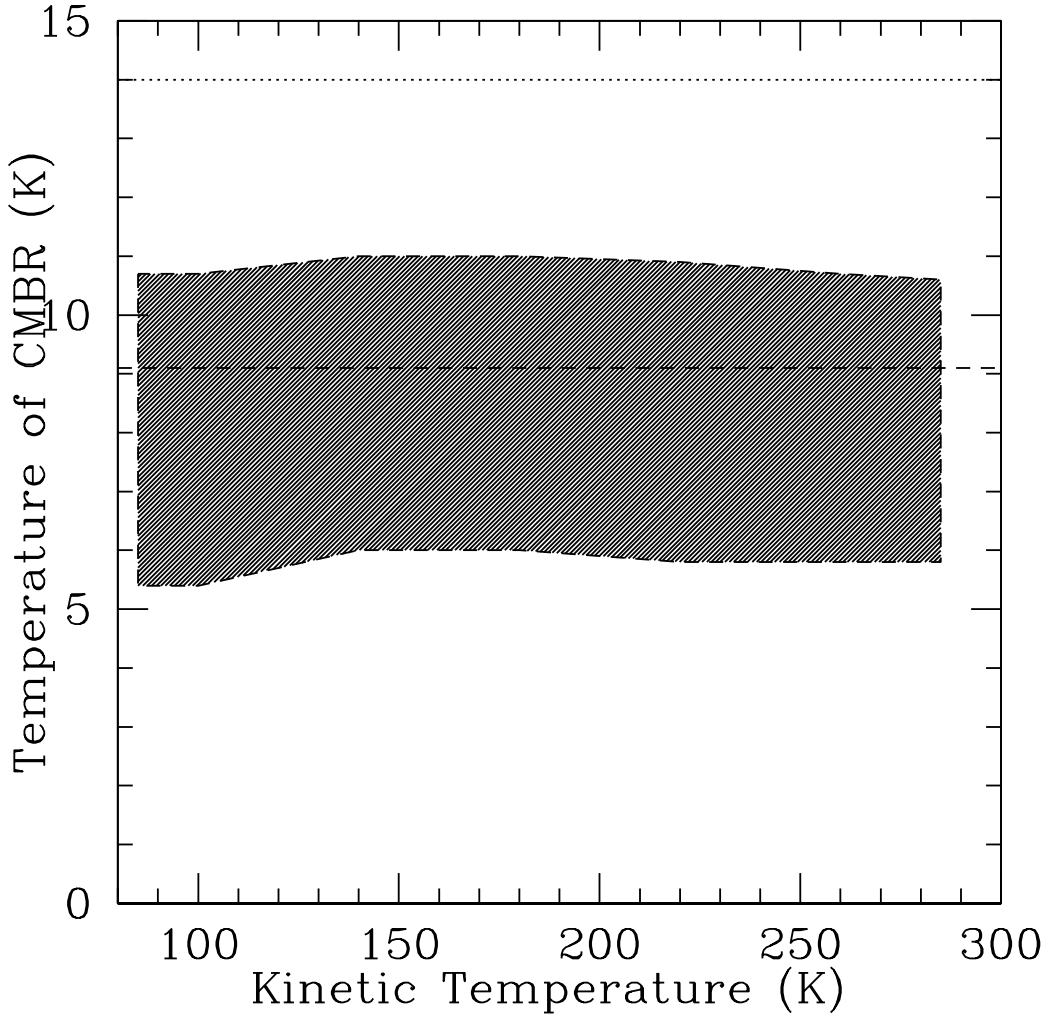,height=14.cm,width=14.cm,angle=0}
}}
\figcaption[cmbfig4.ps]{{\bf Cosmic microwave background temperature as a function
of kinetic temperature of the gas:} 
For a given temperature we obtain an upper limit of the 
hydrogen density from the C$^+$ fine-structure population. 
This is used to derive the contribution of 
hydrogen collisions to the C$^0$ fine-structure excitation.
The excess is used to determine the range for the CMBR temperature.
The shaded region gives the 2$\sigma$ range of the Cosmic Microwave Background
Radiation temperature allowed by the observed population ratios of 
neutral carbon fine-structure levels.  
The horizontal dotted line is the upper limit 
on T$_{\rm CMBR}$ assuming CMBR to be the only 
excitation process. 
The dashed line is the predicted temperature
at z~=~2.33771 in the standard big$-$bang model. This 
demonstrates the presence of a background radiation with
a temperature at least twice that measured in the local universe,
T$_{\rm CMBR,0}$~=~2.726~K.
\label{cmb}}
\end{figure}

\noindent
More generally, we estimate the allowed range for the CMBR temperature
as a function of kinetic temperature, after taking into account all
previously discussed processes.  The allowed area in the T$_{\rm
CMBR}$--T$_{\rm Kinetic}$ plane is shown as the shaded region in
Fig.~\ref{cmb}.  This region is obtained using the upper limit on the
hydrogen density derived above and the 2$\sigma$ ranges of the
C$^{0*}$/C$^0$ and C$^{0**}$/C$^0$ ratios.  Thus the lower T$_{\rm
CMBR}$ boarder of this area gives a stringent lower limit on the CMBR
temperature at $z_{\rm abs}$~=~2.33771.  \par\noindent Note that an
upper limit on T$_{\rm CMBR}$ is obtained assuming that the CMBR is the
only excitation process at work. This limit is shown as a dotted line
in Fig.~\ref{cmb}.
\par
\begin{figure}
\centerline{\vbox{
\psfig{figure=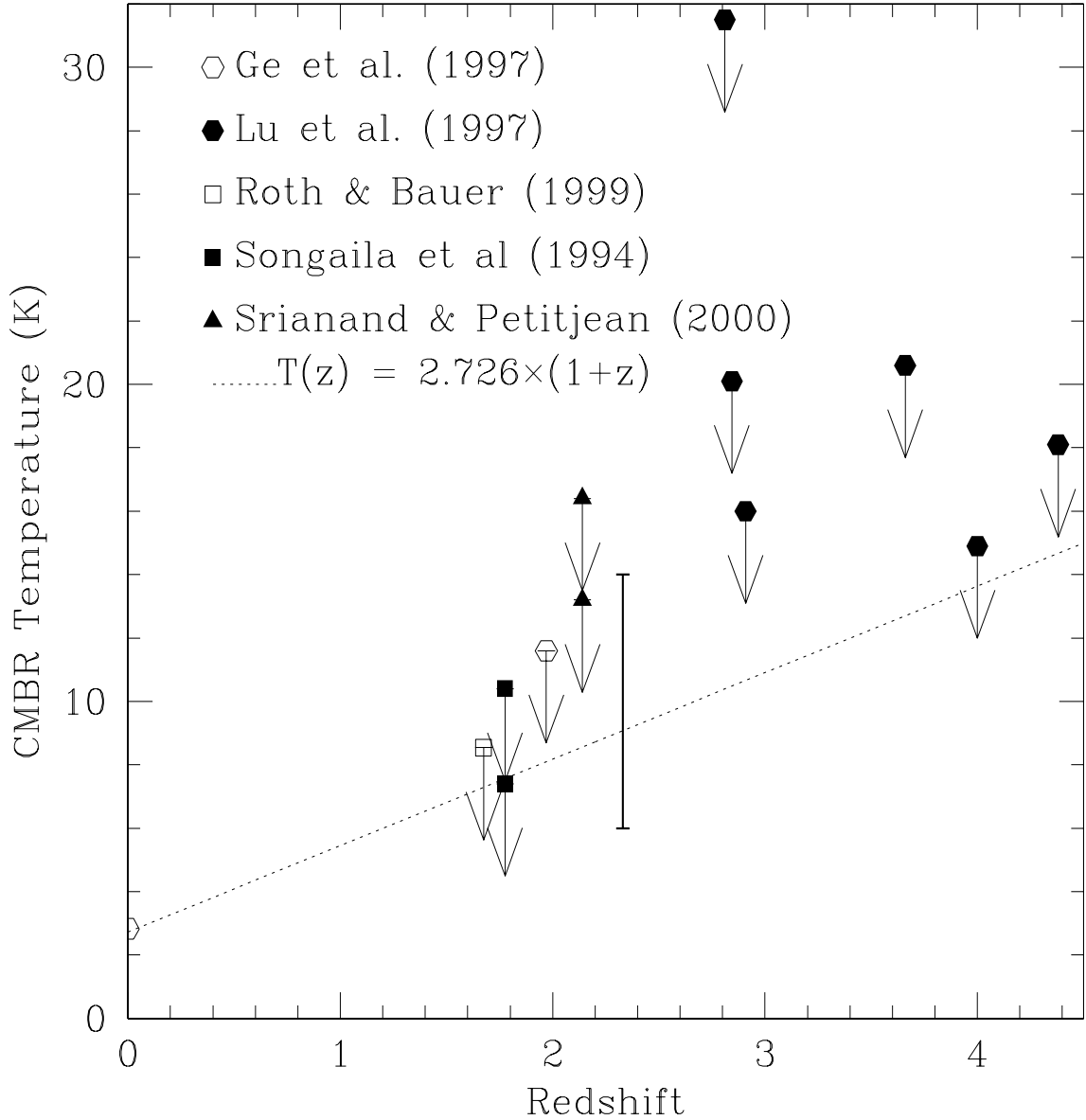,height=17.cm,width=16.cm,angle=0}
}}
\figcaption[]{{\bf Measurements of the Cosmic Microwave Background Radiation 
temperature at various redshifts:}
The point at $z$~=~0 shows the result of the Cosmic Background 
Explorer (COBE) determination$^{\cite{mat}}$, $T_{\rm CMBR}$(0)~=~2.726$\pm$0.010~K.
Upper limits are previous measurements$^{\cite{son1},\cite{Lu}, \cite{ge},
\cite{roth}}$ using the same techniques as
in this paper. We also include our two new unpublished upper limits at
$z~=~2.1394$ along the line of sight toward Tololo 1037$-$270.
The measurement from this work, 6.0~$<$~$T_{\rm CMBR}$~$<$~14.0~K at 
$z$~=~2.33771, is indicated by a vertical bar. The dashed line is the 
prediction from the Hot Big Bang,
$T_{\rm CMBR}$($z$)~=~$T_{\rm CMBR}$(0)$\times$(1~+~$z$). 
\label{tcmbr}}
\end{figure}

\noindent
The presence of a relic radiation field at any redshift is more or less
admitted by most of the practicing cosmologists. Its reality is proved
observationally only in the local universe through direct 
measurements$^{\cite{mat},\cite{roth1}}$
and the Sunayev \& Zeldovich effect$^{\cite {sz}}$. Our
measurement demonstrates for the first time that the cosmic radiation 
exists at earlier times and has a higher temperature than today. 
In the standard Big-Bang cosmology the cosmic microwave
background is a relic radiation left over from an early hot phase.  In
such a model the radiation evolves adiabatically in the expanding
universe and the temperature at any redshift is simply
$T(z)=T(z=0)(1+z)$.  We summarise all the available upper limits  in
Fig.~\ref{tcmbr}, together with our measurement. The prediction for the
adiabatic expansion is  shown as a dotted line.  All the upper limits
and our measurement are consistent with the predictions of the standard
model.
Doing similar analysis over a large redshift range will provide a
direct and model independent measure of the evolution of the cosmic
microwave background radiation.
\par\noindent

\clearpage
\newpage
\par\bigskip\noindent
\clearpage
\newpage
\par\bigskip\noindent
{\bf Acknowledgments}
The observations presented here have been obtained using the Ultra-violet and 
Visible Echelle 
Spectrograph mounted on the 8.2~m KUEYEN telescope operated 
by the European Southern Observatory at Paranal, Chile. 
We gratefully acknowledge support from the Indo-French Centre 
for the Promotion of Advanced Research (Centre Franco-Indien pour la Promotion
de la Recherche Avanc\'ee).
PPJ thanks Andreas 
Kaufer and Merieme Chadid for their kind and efficient assistance at the  
telescope and 
IUCAA for hospitality during the time this work was completed. We 
thank T. Padmanabhan for useful comments.

\clearpage
\newpage


%
\begin{table}
\begin{center}
{\bf TABLE 1: Heavy elements from the damped Lyman-$\alpha$ system}
\\
\begin{tabular}{lcccc}
\\
\multicolumn{1}{c}{Ion}& \multicolumn{1}{c}{log $N$~(cm$^{-2}$)}&
\multicolumn {1}{c}{$b$~(kms$^{-1}$) }&
\multicolumn{1}{c}{[Z/H]} & \multicolumn{1}{c}{[Z/H]$-$[Z/H]$_\odot$}\\
\\
H{$^0$}       & \multicolumn {1}{c} {20.90$\pm$0.10} & ....&....&....\\
C{$^0$}       & 13.86$\pm$0.22 & 1.70$\pm$0.10 & .... & .... \\
C{$^{0*}$}   & 13.43$\pm$0.07 & .... & .... & .... \\
C{$^{0**}$}& 12.63$\pm$0.22 & .... & .... & .... \\
C{$^{+*}$}  & $\le$ 14.00      & .... & .... & .... \\
Mg{$^0$}      & 13.19$\pm$0.09 & ....         && .... \\
Mg{$^+$}     & 15.44$\pm$0.09 & .... &$-$5.46$\pm$0.13 & 
$-$1.04$\pm$0.13\\
Si{$^+$}     & 15.24$\pm$0.11 & .... &$-$5.66$\pm$0.19&
$-$1.20$\pm$0.10\\ 
Fe{$^+$}     & 14.68$\pm$0.08 & .... &$-$6.22$\pm$0.13& 
$-$1.73$\pm$0.13\\
\end{tabular}
\end{center}
\medskip\noindent
\label{tab1}
$N$ is the column density and $b$ the Doppler width of the line;
$b$~=~(2$kT/m$~+~b$_{\rm turb}^2$)$^{1/2}$, where $m$ is the atomic mass,
$T$ the temperature and $b_{\rm turb}$ the characteristic turbulent velocity
of the gas. [Z/H]~=~log~$N$(Z)$-$log~$N$(H~{\sc i}) is the metallicity and 
[Z/H]$-$[Z/H]$_\odot$ the metallicity relative to the value measured in the
Sun. We used the solar metallicities from Savage \& Sembach$^{\cite{savage}}$. 
All wavelengths and oscillator strengths are from Morton$^{\cite{morton}}$
and Welty et al.$^{\cite{welty}}$.  
\end{table}
\clearpage
\newpage
\begin{table}
\begin{center}
{\bf TABLE 2: Fit results to rotational levels
in the vibrational ground state of H$_2$ }
\begin{tabular}{clcc}
\\
\multicolumn {1}{c}{$z_{\rm abs}$}&\multicolumn{1}{c}{Level}& 
\multicolumn{1}{c}{log $N$~(cm$^2$)}&\multicolumn{1}{c}{$b$~(km~s$^{-1}$)}\\
\\
2.33735&J = 0 &$3.60\pm0.34\times10^{15}$&24.07\\
      &J = 1 &$3.34\pm0.40\times10^{15}$&24.07$\pm$10.20\\
2.33754&J = 0 &$4.10\pm0.48\times10^{15}$&14.09\\
      &J = 1 &$2.65\pm0.44\times10^{15}$&14.09$\pm$4.57\\
2.33771&J = 0 &$2.30\pm1.00\times10^{16}$&4.62$\pm$0.36\\
      &J = 1 &$8.30\pm2.60\times10^{16}$&4.62    \\
      &J = 2 &$1.59\pm0.27\times10^{16}$&4.62    \\
      &J = 3 &$9.39\pm1.10\times10^{15}$&4.62    \\
      &J = 4 &$4.42\pm0.80\times10^{14}$&4.62    \\
      &J = 5 &$4.64\pm0.77\times10^{14}$&4.62    \\
2.33793&J = 0 &$1.00\pm0.04\times10^{15}$&21.74\\
      &J = 1 &$6.69\pm0.40\times10^{15}$&21.74$\pm$1.40\\
\end{tabular}
\end{center}
\label{tab2}
\end{table}
\clearpage
\newpage
\vskip -1cm
%
%
%
%
%
%
\newpage
%

%
\end{document}